\documentclass{article} \usepackage{style,times}
\usepackage{amsmath} \usepackage{amsfonts} \usepackage{amssymb}
\usepackage[dvips]{graphicx}
\usepackage{apacite}
\usepackage[english]{babel}

\title{Emotional Reactions and the Pulse of Public Opinion: Measuring
  the Impact of Political Events on the Sentiment of Online
  Discussions}

\author{
Sandra Gonzalez-Bailon\\
Oxford Internet Institute\\
OX1 3JS, Oxford, UK\\
\texttt{sandra.gonzalezbailon@oii.ox.ac.uk} \\
\AND
Rafael E. Banchs \\
Human Language Technology, Institute for Infocomm Research \\
1 Fusionopolis Way, \#21-01, Connexis South, Singapore 138632\\
\texttt{rembanchs@i2r.a-star.edu.sg} \\
\AND
Andreas Kaltenbrunner \\
Barcelona Media – Innovation Centre \\
Av. Diagonal, 177, 08018 Barcelona, Spain\\
\texttt{andreas.kaltenbrunner@barcelonamedia.org} \\
\
}

\begin{document}

\maketitle

\begin{abstract}
  This paper analyses changes in public opinion by tracking political
  discussions in which people voluntarily engage online. Unlike polls
  or surveys, our approach does not elicit opinions but approximates
  what the public thinks by analysing the discussions in which they
  decide to take part. We measure the emotional content of online
  discussions in three dimensions (valence, arousal and dominance),
  paying special attention to deviation around average values, which
  we use as a proxy for disagreement and polarisation. We show that
  this measurement of public opinion helps predict presidential
  approval rates, suggesting that there is a point of connection
  between online discussions (often deemed not representative of the
  overall population) and offline polls. We also show that this
  measurement provides a deeper understanding of the individual
  mechanisms that drive aggregated shifts in public opinion. Our data
  spans a period that includes two US presidential elections, the
  attacks of September 11, and the start of military action in
  Afghanistan and Iraq.
\end{abstract}

\section{Introduction}
Public opinion is a proxy for the way citizens perceive political
issues and react to current affairs. Scandals or controversial
policies, natural disasters or international conflicts, can all
provoke shifts in the opinions of the public and cast shadows over the
authority of their representatives. Public opinion can impact on the
political process by means of electoral accountability, or by means of
propaganda and media manipulation (Glynn, Herbst, O'Keefe, and Shapiro
1999; Jacobs and Shapiro 2000; Lewis 2001; Zaller 1992). Both channels
offer a link connecting the people with their leaders that is central
to the democratic process and to the legitimacy of policy making
(Lippmann 1922). Citizens can use public opinion to articulate their
interests and reward or punish their representatives for their
actions; and political leaders can adapt their discourse and
performance to the interests of their constituents by tracking their
opinions (Delli Carpini and Keeter 1996; Hutchings 2005). Knowing what
the people think and what prompts changes in those opinions is a core
element of democratic governance, and the reason why vast amounts of
efforts and resources are being employed to measure it.

Several barometers are designed to track shifts in public
opinion. Approval ratings, for instance, offer monthly measures of
support to government; and several sample surveys gauge public opinion
around a range of controversial issues like abortion, arms control or
gay rights (Erikson, MacKuen, and Stimson 2002; Stimson 1998; Stimson
2004). While approval rates offer a continuous but shallow measure of
what the public thinks, surveys are richer but usually designed to
capture long-term dynamics on very specific areas of public
concern. In this paper we propose a new approach to the study of
public opinion that aims to complement these previous efforts and move
forward our understanding of how the public thinks. The novelty of our
approach is twofold: we analyse what the public decides to discuss
about, as opposed to their opinions on a battery of predetermined
topics; and we extract and analyse the emotional content of their
discussions on a large scale, for which we use the Affective Norms for
English Language Words (Bradley and Lang 1999; Dodds and Danforth
2009). Emotions have been theorised as a fundamental antecedent of
human action: they affect how individuals process information, but
also how they form their preferences, desires and beliefs (Elster
1999; Frank 1988; Frijda 1986; Turner and Stets 2006). Appraisal and
affective intelligence theories have long considered the cognitive
effects of emotions and their role in public opinion formation
(Lazarus 1991; Lazarus and Lazarus 1994; Marcus, Neuman, and MacKuen
2000). Our approach builds on this line of research, but shifts
attention from the level of individual reaction to the aggregated
patterns of general sentiment. 

This paper aims to show that online communication, in the form of
public discussions, offers new empirical insights into how the public
responds to political events. The main assertion this paper makes is
that online discussions are an untapped domain in political
communication research, and it offers a strategy to start mining that
domain. The analyses that follow make two main contributions: first,
they show that online discussions, although not demographically
representative of the population, are representative of public opinion
trends (as measured by presidential approval rates); and second, that
emotions can be used as consistent indicators of political
attitudes. The paper proceeds as follows. First, it reviews previous
attempts to track public opinion, focusing on findings about the
volatility and polarisation of public views; and it summarises
political psychology research on how emotions mediate information
processing and attentiveness to political events. Then, it presents
the data and methods used to track general sentiment, involving tens
of thousands of discussions held in several Usenet political
newsgroups during a six-year period (1999-2005). Section four shows
that the political events that took place in this period were
accompanied by different emotional responses as measured in terms of
valence, arousal, and dominance; it also shows that changes in these
three emotional dimensions correlate (albeit to a different degree)
with changes in presidential approval ratings. The paper ends with a
discussion of the implications that these findings have for public
opinion research, qualifying arguments about polarisation and opinion
tides, and confirming some of the findings of affective intelligence
theory.

\section{Research on Public Opinion and Emotional Reactions} 
Changes in public opinion follow different frequencies: there is the
slow pace of issue evolution and ideological alignment, which tracks
the distribution of political preferences in the population over
decades of generational shifts (Erikson, MacKuen, and Stimson 2002;
Page and Shapiro 1992); and there is the fast responsiveness of
approval rates, which offer a volatile measure of what the public
thinks (Clarke, Stewart, Ault, and Elliott 2004; Kriner and Schwartz
2009; Mueller 1973). Research on issue alignment and issue evolution
has focused on the changing salience of conflicts over public policy,
and on how the electorate position themselves with regard to those
issues. This includes policies about education, race, welfare or
health care, but also gun control, capital punishment or abortion, all
of which generate public debates that change in intensity and
visibility over the years (Adams 1997; Carmines and Stimson 1989;
Layman 2001; Schuman, Steech, and Bobo 1985; Stimson 2004; Wolbrecht
2000). While most of these issues are not a priority for the vast
majority of the public (whose knowledge about politics is consistently
low anyway, Delli Carpini and Keeter 1996), they are closely monitored
by minorities that turn those issues into decisive electoral factors
and stir awake the ``sleeping giant'', as some have called the general
public (Hutchings 2005). These issues define the electoral boundaries
of public opinion because they open the competition space for
political parties (Carmines and Stimson 1989; Petrocik 1987; Repass
1971; Stokes 1963): parties want to be aligned with what the
electorate think and match the policy expectations of as many voters
as possible. The composition of those expectations, however, changes
over time.

One of those changes is the increasing levels of polarisation bred by
public opinion during the last few years: today there are more people
adopting extreme positions in their policy alignment than decades ago
(Baldassarri and Gelman 2008; DiMaggio, Evans, and Bryson 1996; Evans
2003; Fiorina, Abrams, and Pope 2005; Layman 2001). There is no
consensus about the reach or extent of that polarisation: recent
literature suggests that it affects only the most contended issues
(like, for instance, the war in Iraq), and that the gap between
extremes is actually being widened by political parties, who are
becoming better at sorting individuals along ideological lines
(Baldassarri and Gelman 2008; Fiorina, Abrams, and Pope
2005). Polarisation has not taken place around ``valence'' issues (those
that are uniformly liked or disliked by the electorate, regardless of
their ideological leaning, like for instance corruption, Stokes 1963);
and recent findings show that the basic correlation structure that
links issues into belief systems has in fact remained pretty stable
over time (Baldassarri and Goldberg 2010): the way people think about
some issues (i.e. abortion) is not independent of their opinions about
other issues (i.e. gay rights), and the structure of this
interdependence does not seem to have shifted that much over the last
few decades.  

The debate around polarisation has been accompanied by a parallel
debate on the measurement problems associated to survey
research. Research on public opinion often assumes that people's
preferences are well defined and consistent, and if they are not, that
their inconsistencies are cancelled out on the aggregate level
(Converse 1964; Page and Shapiro 1992; Zaller and Feldman 1992; Zaller
1992). This assumption derives from the way opinions are elicited,
namely using a battery of issues (primed by the media, the political
discourse, or the researcher) on which respondents are asked to give
their views. This measurement strategy generates the problem of
non-attitudes: respondents might not have an opinion formed around the
issues they are being asked about; it also ignores a wide range of
topics in which the public might be more interested. Surveys using the
open ended ``most important issue'' question, on the other hand, have
other measurement problems, mostly their inability to differentiate
the importance of an issue from the degree to which issues are a
problem (Wlezien 2005). This means that research on public opinion
that relies on those surveys have some intrinsic limitations in
what they can say about the origins and shifts of mass opinion.

Research on approval rates puts together the opinion of the public
into a single measurement: their satisfaction with the management of
government. This approximation to what the public thinks is not flawed
by the problems identified above because it does not go into the
reasons why respondents approve (or not) the President's job. On the
aggregate, this measurement is rich enough to help identify inertias
that systematically appear during the life cycles of all
administrations. Previous research has identified consistent
``honeymoon'' periods during the first months in office, when there is
a general level of contentment and approval rates are higher; it has
also identified an unavoidable attrition paired with the act of
governing, which makes falling rates just a matter of time. In times
of crisis, approval surges systematically, in support of leaders, and
it tends to move in parallel with the national economy: when
prosperous, rates are higher, but when the economy goes down, so do
approval rates. Finally, research on approval rates shows that they
also tend to equilibrate over time: when they are above or below 50
percent, they tend to converge back to it (Stimson, 2004: 144-148;
also Mueller, 1973). Controlling for these trends, the rest of the
variation in approval depends on particular political events, like the
Watergate scandal in the 70s, or the Lewinsky case in the 90s. This
line of research is based on the assumption that the public is well
informed, that they follow closely political events and react to them
as a thermostat reacts to room temperature.

These two areas of research, changes in issue alignment and approval
rates, support the old claim that politics is as much about passions
as about reasons: on the one hand, people choose which controversies
they feel excited about, channelling their political engagement around
those issues (Converse 1964: 245); on the other hand, the evaluation
that the public makes of their representatives reflects more the
general sentiment of hope or despair than the actual performance of
government (Stimson 2004: 154). Recent developments in political
psychology show increasing efforts to identify the role that emotions
play in the formation of political choices (Mutz 2007). Most of these
efforts, including the theory of affective intelligence, revolve
around the cognitive effects of emotions, and how they act as
heuristic devices that voters use to gather and process information
(Marcus and MacKuen 1993; Marcus, Neuman, and MacKuen 2000; Neuman,
Marcus, Crigler, and MacKuen 2007; Rahn 2000; Sniderman, Brody, and
Tetlock 1991). One of the main findings of affective intelligence
theory is that anxiety about political affairs makes people become
more alert and vigilant, and more inclined to gather and process
relevant information; in other words, this negative emotion makes
citizens more thoughtful and attentive (Marcus, Neuman, and MacKuen
2000). A more recent study has also shown evidence of the effects that
anxiety and anger have on political attitudes in the context of the
Iraq war.Both emotions increased attention to the war, but they had
opposite effects on support: anger increased support to the military
intervention, but anxiety reduced it (Huddy, Feldman, and Cassese
2007). One acknowledged weakness of these studies is that they rely
too heavily on induced emotions and retrospective self-report, which
makes them vulnerable to measurement errors, and difficult to
generalise (Mutz 2007). This paper aims to overcome these weaknesses
by proposing an alternative way to measure emotional reactions to
political events like the war in Iraq.

Our analytical strategy consists on analysing real-time reactions to
political events (i.e. as they happen, not as they are recalled) by
measuring and aggregating the emotional content of opinions
voluntarily expressed over time and on a large scale. This adds what
we think is an unexplored dimension to the analysis of emotions and
political behaviour: we know very little about which emotions prevail
on the population level over time, or how long it takes for the
effects of emotions to decay. The political psychology literature
differentiates between ``mood'' and ``emotional reactions'' because
the former refers to a much longer phenomenon than the latter, which
is more focused and short-lived. By tracking general sentiment over
time, our approach gives an empirical criterion to assess when
emotional reactions crystallise into mood; it also offers a glimpse
into the ``dark matter'' or unseen forces that operate in the
background of public opinion (Stimson 2004: 144). Our approach,
however, should be seen as complementary to other political psychology
approaches: a topic as elusive and complex as the role of emotions in
public opinion formation needs to sum, rather than subtract, sources
of data. Enriching this essential toolbox is, ultimately, our goal.

There are two types of questions that motivate this research. The
first, methodological, is can we solve some of the problems associated
to survey research by implementing a bottom-up approach to the study
of public opinion? By bottom-up we mean analysing the opinion
voluntarily expressed by people in politically relevant
discussions. For that, we propose exploiting the new forms of
interpersonal communication enabled by Internet technologies. We have
considered elsewhere whether online interactions can promote
deliberation and encourage the construction of politically relevant
discussion networks (Gonzalez-Bailon, Kaltenbrunner and Banchs 2010;
Gonzalez-Bailon 2010). The purpose here is to determine
whether the opinions expressed in those discussions can be used to
assess emotional reactions to political events, and whether those
reactions co-evolve with political attitudes. The second type of
questions are theoretical: First, what can the analysis of general
sentiment add to the polarisation debate? Does the way people react to
political events suggest convergence or polarisation? Emotions are
universal mechanisms, but the same stimuli might generate
heterogeneous reactions; our aim is to find out how much variance
different events generate. And second, what emotional dimension
explains better political attitudes? Affective intelligence theory
suggests that anxiety and anger have vigilant effects on citizens:
they become more alert the more threatened they feel. Do our data
support the prevalence of this type of emotion, and its higher
predictive power of political attitudes? The analyses that follow are
an attempt to answer these questions.

\section{Methods and Data: Measuring Emotion in Online Discussions}
The data we use tracks political discussions in the online forum
Usenet, a distributed discussion system that has been active for over
three decades (Hauben and Hauben 1997; Lueg and Fisher 2003). We use
the dataset Nestscan (Smith 2003; Smith and Kollock 1999), a sample of
Usenet that contains about 350 thousand discussion groups. Our
analyses focus on the discussions held within the groups that
contained the word ``politics'' in their hierarchy (hierarchies are used
to organise newsgroups in nested categories); for the period September
1999 to February 2005, the time window we analyse, this totalled 935
groups, involving about 800 thousand unique participants. We chose to
analyse political discussions in Usenet because they allow us to
reconstruct patterns over a longer time period than more recent social
media. The time window considered here includes some prominent events
like two presidential elections, the attacks of 9/11 or the invasion
of Iraq. The series we analyse are aggregated on a monthly basis. When
preparing the data for the analyses, we excluded the threads that did
not have at least three messages in order to avoid spam and
non-significant discussions.  

The discussions we track with this data involve users that are
obviously not a representative sample of the population:
representativeness is undermined not only by the digital divide
(particularly important towards the beginning of the period, when
Internet penetration rates were lower) but also by the self-selecting
nature of these groups. Our discussants are probably more interested
in politics than the average person: after all, we analyse the
discussions of a minority (of hundreds of thousands) of users
sufficiently engaged in politics to be active in these
forums. However, this phenomenon is not specific to online data:
political engagement is always the inclination of a few, and those
that usually broadcast the signal of public opinion are a minority as
well (Hutchings 2005). People who participate in the discussions we
analyse are the equivalent to opinion leaders, or to the ``passionate
people'' to which the literature on public opinion usually refers
(Stimson 2004: 163): they are those who care a great deal about public
affairs and have strong enough views to talk about them. If we assume
that around this core of active discussants there is an even larger
periphery of users that follow the discussions without participating,
then the role of these opinion leaders in defining general sentiment
becomes even more important. The signal of public opinion that we
capture with this data is, nonetheless, strong and diverse: the
discussion groups we track are positioned along the ideological scale,
which can be assessed using the names of the discussion groups
(i.e. alt.politics.democrat, alt.politics.republican).

We measure public opinion using the messages that users contribute to
these forums. In particular, we analyse the emotional content of the
discussions following the method proposed by Dodds and Danforth
(2009). This method employs the ANEW (Affective Norms for English
Language Words) list to quantify the emotional content of texts on a
continuous scale. The ANEW list contains about a thousand words that
receive a rating on a 9 point scale in three dimensions: valence,
arousal, and dominance (Bradley and Lang 1999). The valence dimension
associates words with feelings of happiness, pleasure, satisfaction or
hope: the higher the score a word receives in this dimension, the
happier it makes subjects feel, on average. The arousal dimension
measures the extent to which words make subjects feel stimulated,
excited, or frenzied as opposed to calm, dull, or sleepy: again, the
higher the score of a word in this dimension, the more aroused it
makes subjects feel. Finally, dominance focuses on feelings of
domination or being in control versus feelings of submission or awe;
the higher the score of a word in this dimension, the more it is
associated with feelings of autonomy and prominence. On the grounds of
the literature reviewed in the previous section, our assumption here
is that the emotional response of citizens to political events
precedes their rationalisation of opinions and attitudes.  

There are other examples of large-scale text analysis that use
alternative methods to gauge emotions or sentiment in public
opinion. These examples use micro-blogging communication (i.e. Twitter
messages) to measure what people think about politics (O'Connor,
Balasubramanyan, Routledge, and Smith 2010) and movies (Asur and
Huberman 2010), and use those opinions to make market-based
predictions in the form of Box-office revenues and presidential
popularity scores. We believe that our approach improves on these
efforts not only because it covers a wider time window which, as
mentioned above, helps us identify long-term shifts in public opinion;
but also because it is based on a lexicon of emotional words that has
been validated by psychologists (as noted by Dodds and Danforth 2009)
and replicated in other languages (i.e. Redondo, Fraga, Padron, and
Comesana 2007), which potentially widens the comparative dimension of
the analyses across linguistic communities. We agree with these other
approaches in that using online data is faster, less expensive and,
crucially, more informative in important ways than traditional surveys
or polls.

\begin{figure}[!tb]
\begin{center}
  \includegraphics[angle=0,width=\textwidth]{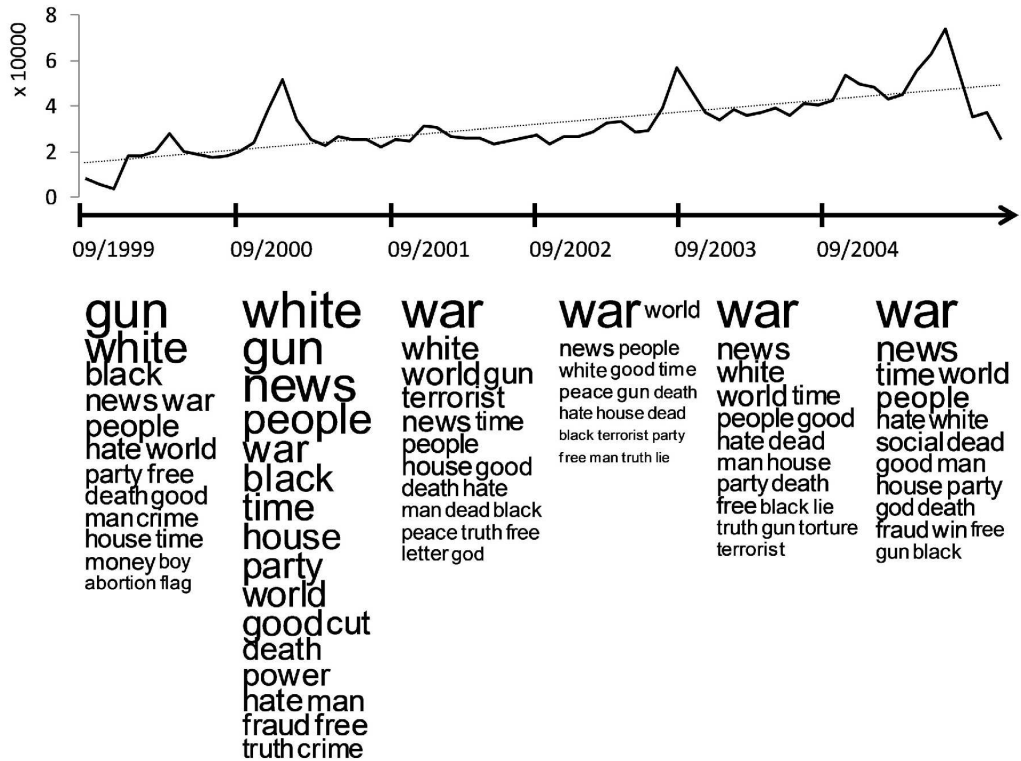}
\end{center}
\caption{Number of discussions and most popular ANEW words used over time.}
\label{fig:1}
\end{figure}

In our analysis, we use the subject line of the messages as a proxy to
the topic of the discussions as well as a summary of their contents;
for instance, one popular discussion had the subject line ``Bigots for
Bush'' (with about 19,000 messages and 690 unique discussants), and
one unpopular discussion had the subject line ``John Edwards, weak but
has nice hair'' (with only 3 messages sent by 3 different
discussants). In total, we analysed about 380 thousand subject lines
containing about 2,3 million words, 6\% of which (N~140,000) appeared
in the ANEW list. The low percentage of words contained in the list
means that this analytical strategy is only reliable when large
corpuses of text are available (Dodds and Danforth 2009). Figure 1
contains the list of twenty most popular ANEW words aggregated per
year in the form of tag clouds (i.e. the size of the word corresponds
to the square root of its number of occurrences in the subject
lines). The series at the top of the figure indicates the total number
of discussions initiated on a monthly basis; the ANEW words at the
bottom are sized in proportion to their relative weight within the
discussions held every year, starting in September 1999. The figure
shows two things: first, that there is an upward trend in number of
political discussions initiated in these forums: towards the end of
the period there are twice as many discussions around political issues
than at the beginning; and second, that there is a shift in the
visibility of certain topics. After 9/11, ``war'' becomes the most
prominent topic, clearly outweighing the other topics in the
discussions held between September 2003 and September 2004, the year
of the invasion of Iraq.


Every word in the ANEW list has an average score in the three scales
(valence, arousal, and dominance). Following Dodds and Danforth (2009)
we counted the number of instances in which ANEW words appear in the
subject lines of the discussions, and calculated monthly averages and
standard deviations based on their frequency and scores. However,
unlike Dodds and Danforth we not only analyse valence values, but also
those in the arousal and dominance dimensions. As the previous section
showed, these three dimensions represent different psychological
mechanisms, and prompt different behaviour (Marcus, Neuman, and
MacKuen 2000; Neuman, Marcus, Crigler, and MacKuen 2007). Which of the
three emotional dimensions (valence, arousal or dominance) explains
better political attitudes is one of the substantive questions we want
to answer with this data. The second question is how volatile
emotional responses are, and how much disagreement or polarisation
there is around average values. If emotions play a substantive role in
shaping preferences and behaviour, as affective intelligence theory
suggests, the heterogeneity of their distribution is also an important
element when it comes to inferring political outcomes: if the
population grows increasingly angrier, one should expect more informed
and vigilant citizens.
 
\section{Shifts in General Sentiment and Approval Rates}
\subsection{The Impact of Political Events on Emotional Reactions}
This section pays attention to shifts in emotional reactions as
measured by changes in the three dimensions: valence, arousal, and
dominance. Figure 2 shows the emotional scores on these three
dimensions, plotting averages (left y-axis), and standard deviations
(right y-axis) in the same figure as they change over time. The grey
vertical bars identify some of the most prominent events in this
period: the two presidential elections (in November 2000 and November
2004), the attacks of September 11, the invasion of Iraq, and the
abuses of Abu Ghraib. These series show two clear trends: first, that
9/11 marks a before and after, prompting a fall in the average values
of valence and, to a lesser extent, dominance, and a rise in the
scores of arousal; and, second the upheaval caused by the invasion of
Iraq and the scandals associated to the conflict: these events are
associated to the lowest peaks in valence and the highest peaks in
arousal. After 9/11, the standard deviation around mean valence scores
goes up significantly, signalling increasing levels of emotional
polarisation, that is, a higher proportion of messages that fall
closer to either of the two extremes of the happy-unhappy scale.

\begin{figure}[!tb]
\begin{center}
  \includegraphics[angle=0,width=\textwidth]{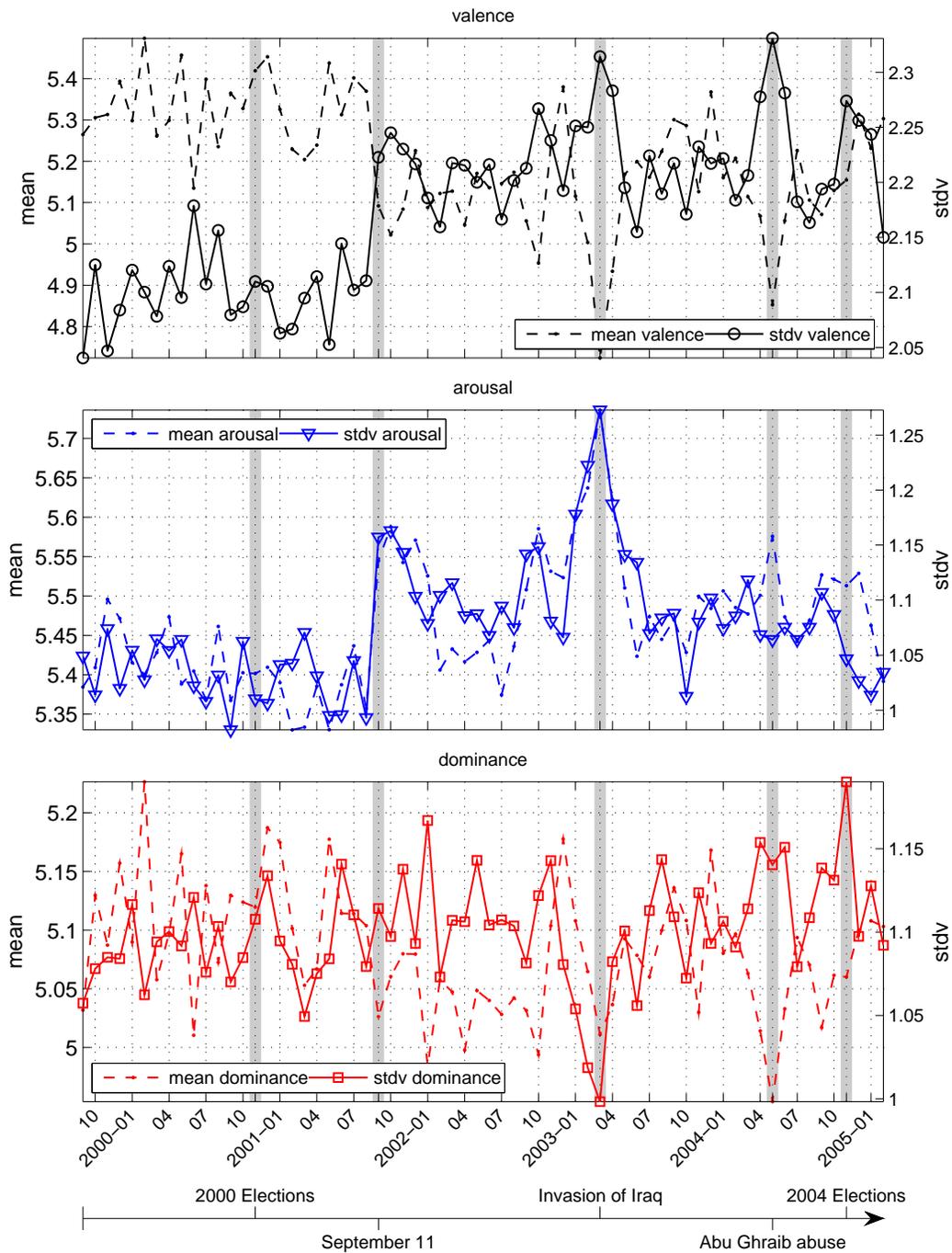}
\end{center}
\caption{Changes in mean and standard deviation of valence, arousal
  and dominance scores over time.}
\label{fig:2}
\end{figure}

Before the attacks, discussions reflected higher levels of happiness
and lower levels of excitement, right the opposite of what happens
after the attacks; the two general elections that took place during
this period did not provoke reactions comparable to those associated
with the military interventions that followed the attack. These trends
show that after 9/11 the public became gloomier and more irascible,
but also that they had more diverse reactions around the average
values. One of the predictions made by political psychology research
is that anxiety and threats make people more alert: risk boosts the
attention of the public. The rise of arousal levels after the attack
indicates that this precondition for a stronger public vigilance is in
place. What these trends add to previous research is the evidence that
changes in emotional reactions still linger years after the stimuli
that originated them took place. This suggests that emotional
reactions to very specific and completely unexpected events can derive
into the more permanent phenomenon of shifting public mood. The
distinction between emotions and mood is important because the latter
permeates the public's perception of events, and affects their
behaviour, for a longer period of time, even after the event that
originated the initial reaction disappears down the time line.  The
political communication tracked in these online discussions shows that
emotions are very volatile, but also that they crystallise into
long-term mood.These time dynamics shed interesting insights into how
permanent emotions shifts are and how long they take to decay.

\subsection{Relationship with Presidential Approval Rates}
The trends showed in Figure 3 reproduce the same information but this
time in comparison to changes in presidential approval ratings, as
obtained from the Gallup organisation's website (for better
visualisation, the standard deviation of valence has been offset by
-1). The figure makes more explicit the relationship between the three
series of emotional reactions and how these are related to changes in
approval rates. As expected, approval reacts sharply to critical
events, noticeably the attacks of 9/11, when support for the President
reached a historical maximum, and the start of military action in
Iraq. Yet when compared to the emotional scores of the discussions,
interesting differences emerge. First, the sudden soar of approval
rates brought about by 9/11 converges, a few months later, back to the
equilibrium of about 50 per cent (in line with the equilibration cycle
of approval rates identified by previous research, Stimson 2004: 145);
yet the average emotional scores do not go back to normal after the
event: as mentioned in the previous section, valence levels remain
lower, and arousal levels higher, than before. This also applies to
the levels of polarisation around valence: they not only remain
higher, but keep on being slightly on the rise.

The shifts caused by this exogenous (and unpredictable) shock take
place at the same time, causing a simultaneous reaction in the four
series. However, different relationships emerge in other points of the
period. The start of the Iraq war, for instance, generates highs and
lows in the three emotional dimensions before it generates a response
in approval rates, which go up shortly after military action starts
(again, in line with what we would expect in times of war, Mueller
1973).  This military intervention coincides with the lowest point in
the valence series, that is, the moment with the unhappiest general
sentiment; it is also one of the moments with the highest emotional
polarisation. The invasion of Iraq also coincides with a peak in
arousal: when the war started, discussions adopted the angriest
expressions of the period considered here. If we just measure public
opinion using approval rates, this war did not bring such an extreme
reaction in the public as the attack of 9/11 had done; but it
definitively stirred more antagonistic feelings. One reason for this
different reaction has probably to do with the unexpectedness of the
attack. The possibility of a war was in the mind of the public for a
longer period, and this gave them more time to digest the news and
have a response ready when the war finally started; in other words,
the public had more time to distil reasons out of their
passions. These findings also qualify previous approaches to the
emotional reactions associated to the war (Huddy, Feldman, and Cassese
2007): anger might increase support, and Figure 3 suggests that the
average feeling was that of an increased anger; but it also shows that
there was quite a lot of divergence around that general feeling, at
least as expressed in written opinion.
 
\begin{figure}[!tb]
\begin{center}
  \includegraphics[angle=-90,width=\textwidth]{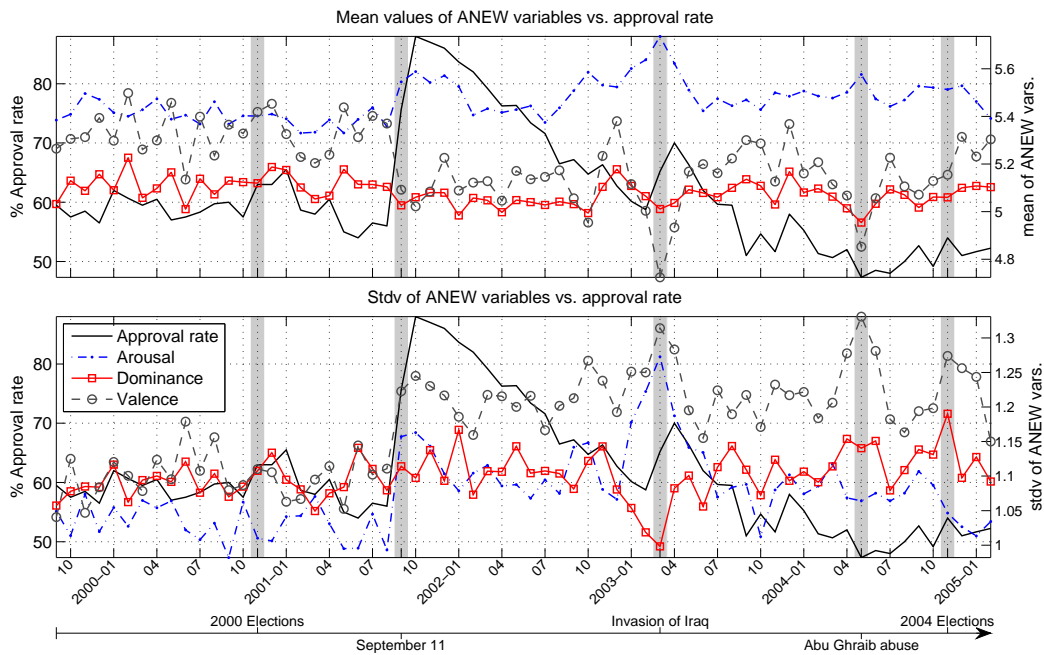}
\end{center}
\caption{Changes in valence, arousal and dominance compared to
  presidential approval rates.}
\label{fig:3}
\end{figure}

Compared to approval rates, these trends suggest that people increase
or decrease their support to the President partly as a result of
different emotional reactions. To identify the degree of association
between the three emotional series and approval rates, we analysed
monthly correlations, plotted in Figure 4. The coefficients show the
degree of association between variables using 13-month time windows,
centred in each corresponding month, to which we assign the
coefficient. At the beginning and at the end of the period we truncate
the time window so that for the first and last month the correlation
coefficient is based on 7 months, for the second and second from the
last, it is based on 8 months, and so forth. From month 7 to month 60
in our series, the correlation is based on the 13-month time
windows. The red intervals depicted in Figure 4 correspond to the
coefficients that proved statistically significant (according to
Fisher's $p$-value with $p<0.05$, Fisher 1925). The first thing the figure
shows is that the three emotional dimensions are not fully
synchronised; the average values of valence and dominance are, for
most of the period, highly correlated, with rates of change that go in
the same direction: the happier the general feeling, the stronger is
also the general sentiment of being in control; but changes in arousal
only correlate sporadically with changes in the other two
dimensions. The strongest correlation with the arousal series is the
standard deviation of valence: excluding the early and late months of
the period, as the standard deviation in valence goes up (that is, as
polarisation increases), the average values of arousal also go up.

The second thing that Figure 4 shows is that the three emotional
dimensions do not correlate to the same extent with presidential
approval rates, shown in the last column of the matrix. While valence
has a significant negative correlation around the period surrounding
the attacks of 9/11 (a correlation that is positive with the standard
deviation series) and a positive correlation during the scandals of
Abu Ghraib, arousal shows the opposite relationship: around the time
of 9/11, higher arousal levels coincide with higher approval rates,
but the relationship is reversed by the time the Iraq invasion
started. Of all three emotional series, dominance is the dimension
that shows the weakest relationship with approval rates. 

\begin{figure}[!tb]
\begin{center}
  \includegraphics[angle=-90,width=\textwidth]{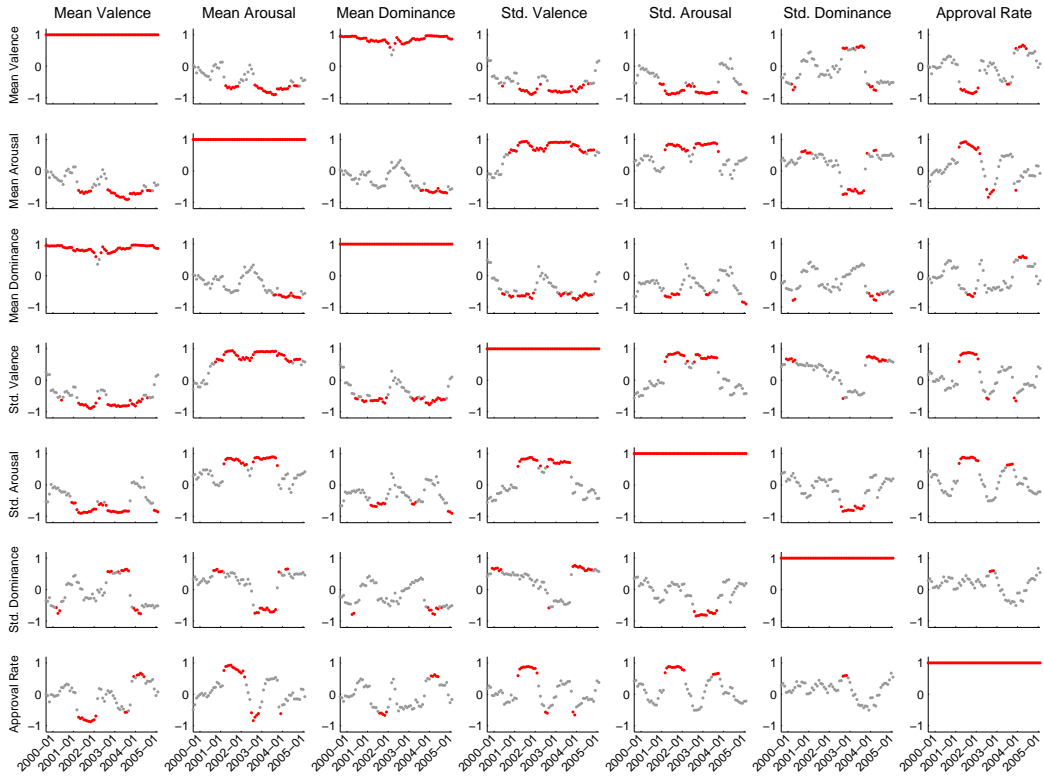}
\end{center}
\caption{Monthly correlations between the series (statistically
  significant intervals are indicated in red).}
\label{fig:4}
\end{figure}

\section{The Predictive Power of Emotions}
The analyses in the previous section show that some of the changes in
approval rates seem to be the consequence of different emotional
reactions, and that some of those emotions have a stronger
relationship with the opinion expressed in the form of evaluations of
the President. This section specifically addresses the question of
which of the three dimensions (valence, arousal or dominance) explains
better changes in presidential approval rates; in other words, the
analyses that follow aim to identify if the emotional content of
political discussions help predict presidential approval rates. Figure
4 showed that the three emotional dimensions do not correlate to the
same extent with presidential approval rates; however, some of the
most relevant variations in their trends occur simultaneously or
closely in time, and seem to be related to salient political
events. To better exploit the dependences among these variables, we
move from linear correlations to a more complex analysis that takes
into account the temporal dimension of the data. In this section we
use time series analysis methods (Box, Jenkins, and Reinsel 2008) to
predict presidential approval rates using the three emotional
dimensions.

We use an ARMA (Auto Regressive Moving Average) model, which takes
into account the recent history of both the response and explanatory
variables (approval rates and the three emotional dimensions) for
predicting approval rates. As the name implies, an ARMA model
incorporates two fundamental components: an Auto Regressive part,
which is able to exploit relevant information related to the
auto-correlated nature of the time series we want to predict
(i.e. presidential approval rates), and a Moving Average component,
which is able to incorporate additional information from other
external time series (in our case, the three emotional
dimensions). The general form of ARMA estimator can be expressed as
follows:
\begin{equation}
X(t)=\sum_{i=1}^p A_iX(t-i)+\sum_{j=1}^m \sum_{i=1}^q B_{i,j}Y_j(t-i)
\end{equation}

where $X(t)$ is the time series to be predicted (i.e. approval rates), $p$
and $q$ are the orders of the auto regressive and moving average models,
respectively (their calibration is explained below), $m$ is the total
number of external sources of information (i.e. the three emotional
series), $Y_j(t-1)$, $t$ is the time index for each time series (i.e. the
66-month measurements), and $A_i$ and $B_{i,j}$ are the model coefficients
that have to be computed during a training phase by using historical
data. The aim of this model is to estimate if changes in the three
emotional time series are significant predictors of changes in
presidential approval rates, once we control for auto-regression, that
is, for the fact that approval rates in time $t$ depend on their value
in $t-1$.  

ARMA models are trained by means of an optimisation procedure aiming
at minimising the fitting error over the training dataset. In our
case, we used the available 66-month data as the training dataset and
evaluated the performance of the prediction models in terms of the
mean absolute error over the same period of time. We use the current
value of the presidential approval rate (AR component) and the 3-month
previous history of the emotional dimensions (MA component) to
generate predictions of the approval rate for the following month
(that is, for $t+1$). According to this, the orders of our ARMA model
are $q=3$ and $p=1$. Prior to running the analyses, all time series
were smoothed using a 4-month window, weighted following a Hamming
distribution to minimise the effects of noise. Smoothing time series
not only allows us to reduce the amount of noise in the data; it also
yields a more robust performance of the prediction
algorithms. Smoothing constitutes a common practice in this kind of
analysis (e.g. O'Connor et al. 2010). A figure comparing the smoothed
and unsmoothed versions of the series can be found in the Appendix.

In a first stage, we constructed and evaluated 10 different
models. The first model, which we use as a benchmark, only
incorporates the autoregressive (AR) component. This model only uses
the information of the approval rate at the current month ($t$) to
estimate the rate at the following month ($t+1$). The model is used to
compare the performance of the other nine models, which incorporate
the moving average (MA) component and were constructed as follows: six
of them include only one emotional time series ($m=1$), either the
mean or the standard deviation of the three emotional variables
(mean-valence, mean-arousal, mean-dominance, std-valence, std-arousal
and std-dominance); the remaining three models include both the mean
and the standard deviation time series ($m=2$) of one emotional
variable at a time (both-valence, both-arousal and
both-dominance). Figure~5 presents mean absolute errors and
accumulated errors for the nine models.

\begin{figure}[!tb]
\begin{center}
  \includegraphics[angle=0,width=0.9\textwidth]{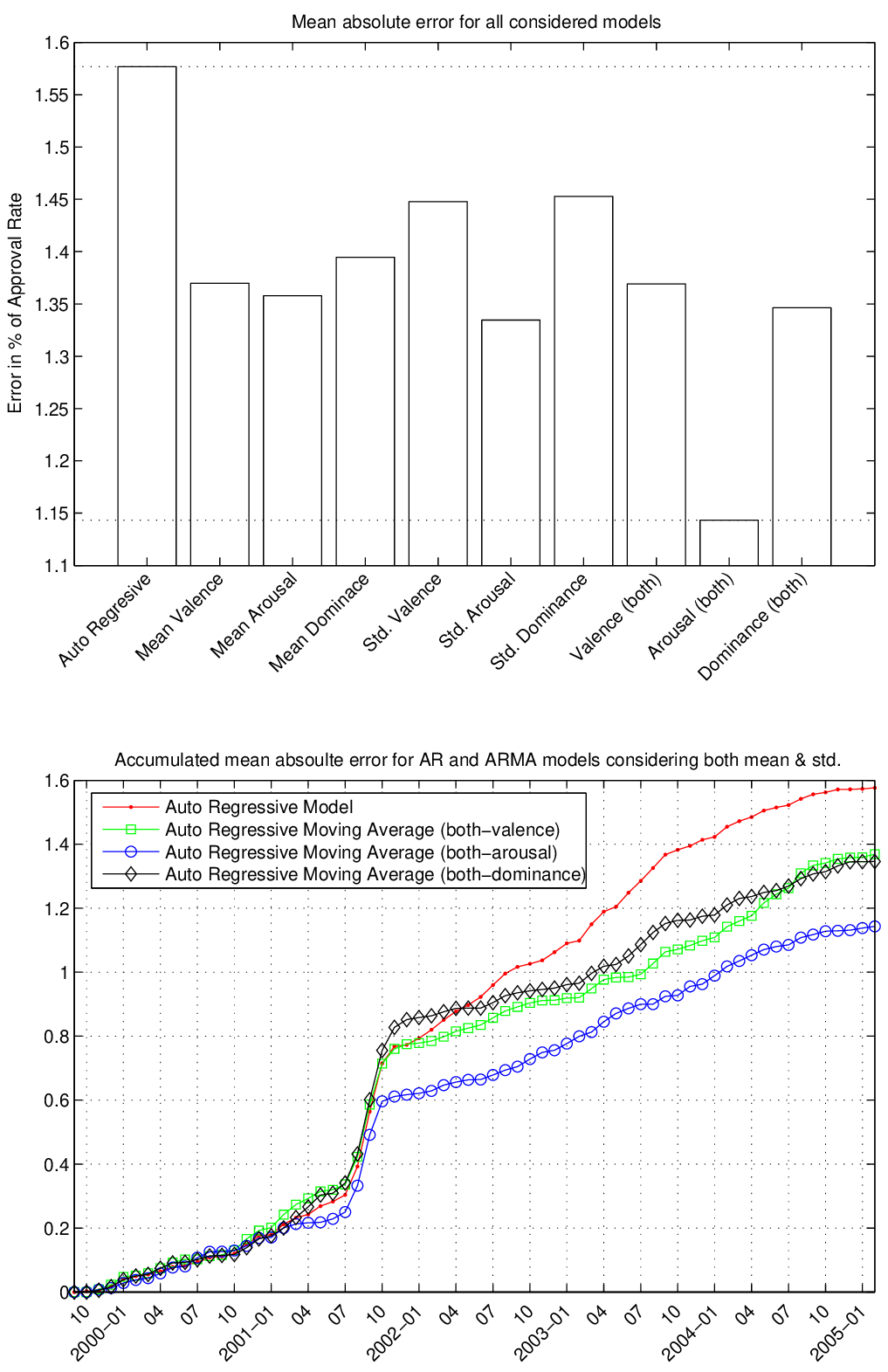}
\end{center}
\caption{Mean absolute errors and accumulated errors for the ten
  predictive models.}
\label{fig:5}
\end{figure}

The figure shows that all ARMA models perform better than the
benchmark AR model, which suggests that the information derived from
the three emotional dimensions increases predictive power. Figure 5
also shows that of all the models that only use one emotional
variable, those considering the mean values generate better
predictions (with the exception of the model with the standard
deviation of arousal). The predictions obtained by the models that
consider both the mean and the standard deviation are not necessarily
better, again with the exception of arousal. Actually, the performance
of the models that take into account arousal-related time series are
always better than those considering valence and dominance.  

The best model in terms of predictive power considers both the mean
and the standard deviation of arousal. The prediction of this model
produces the minimum mean absolute error (MAE=1.14\% of the approval
rate), which is 27\% lower than the benchmark model that only takes
into account the autoregressive component of presidential approval
rates (for which the MAE=1.57\%). To assess the significance of this
difference, we simulated a thousand ARMA models with random inputs
that reproduced the same distribution of the arousal series. None of
these models was able to produce a better prediction than the
empirical ARMA model, which means that arousal variables do help us
predict approval rates, and that this is not just an artefact of
introducing more parameters in the model.  

The detailed time evolution of the prediction errors are presented in
the lower panel of Figure~5. The panel displays the prediction error
for the benchmark AR model and the three ARMA models that consider
both the mean and standard deviation of the emotional variables. These
lines show the cumulative sum of the absolute values of the prediction
errors, normalised by the number of evaluation points (i.e. the last
points in the cumulative curves correspond to the mean absolute error
displayed in the upper panel). Two important observations can be drawn
from this plot: first, the significant contribution of the attacks of
9/11 to the accumulated prediction error for all prediction models
(related to the unexpectedness of the event); and, second, the
consistently better performance of the arousal-related model for the
whole time period.  

In a second stage, we conducted a more detailed comparison between the
benchmark AR model and the best ARMA model, which includes the two
arousal time series. We compared the approval rates predicted by the
two models with the actual approval rates (Figure 6, upper panel), and
their corresponding estimation error curves (Figure 6, lower panel).

\begin{figure}[!tb]
\begin{center}
\includegraphics[angle=-90,width=\textwidth]{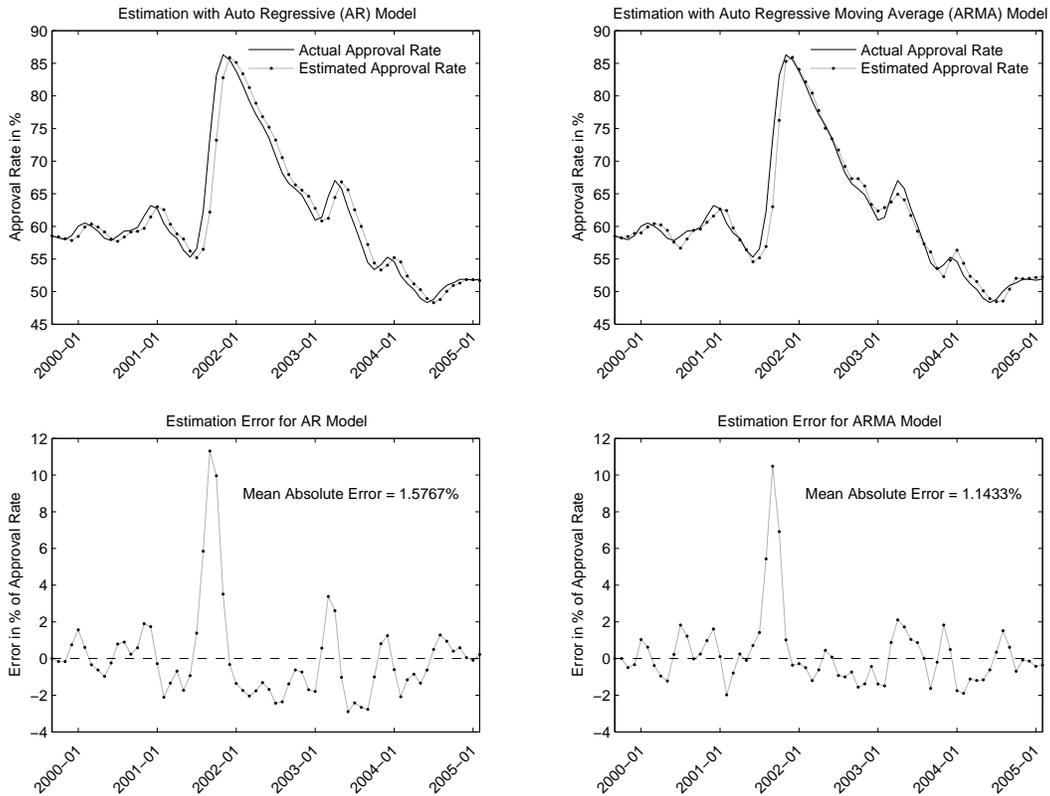}
\end{center}
\caption{Predictions for approval rate with AR and best ARMA models
  and their corresponding estimation error curves.}
\label{fig:6}
\end{figure}

The predictions given by the benchmark model respond to a simple
one-month shift strategy. The autoregressive coefficient obtained by
this model is 0.989, which means that the best guess the model can
make for the next-moth approval rate value is that it is about 0.99
times the current month value, revealing a global downward trend of
approval rates. However, the important point revealed by Figure 6 is
that the ARMA model is able to follow approval rate variations better
than the AR model. The absolute errors curves presented in the lower
panel of Figure 6 show that the largest peak in prediction errors
occurs on 9/11 for the two models. In both cases, the peak is very
similar in amplitude, showing that at this particular point in time
both models are similarly unable to predict approval rates one month
ahead. Again, this is due to the unexpectedness of this event: the
attacks came as an exogenous shock that produced strong reactions in
approval rates and emotional variables (as Figures 2 and 3 showed) but
that is completely unpredictable from the information and previous
history contained in the time series. Leaving this particular case
aside, the figure shows that the information provided by the two
arousal series (mean and standard deviation) certainly contributes to
improve approval rate predictions: while the AR model prediction error
ranges between -3\% and 4\%, the ARMA model's error is bounded between
-2\% and 2\%.

\section{Discussion and Conclusions} 
Can emotions, and their role in public opinion formation, be tracked
using online communication? In the light of our results, the short
answer is yes. We have shown that different political events incite
different emotional reactions, and that these reactions shape
political attitudes, here measured by means of presidential approval
rates. Our findings suggest that arousal is the prevalent emotional
dimension in shaping political attitudes: previous research on
political psychology suggests that a political reality perceived as
riskier and more threatening engenders emotions of arousal; we have
shown that this emotional dimension is also the most significant in
shaping attitudes over time. In line with previous research on public
opinion, we have found evidence of increasing polarisation, at least
since the 9/11 attacks, an event that marked a new era in the cycles
of public opinion and, as our findings suggest, in the mood of the
public. Our findings also show that there are nuances in the
relationship between emotions and public opinion. The same political
events unleash more disagreement in some emotional dimensions
(valence, arousal) than in others (dominance), and they have different
lasting effects: while approval rates tend to equilibrate in the long
run, shifts in the emotional reactions of the public are more
resistant to the weight of time. Although arousal predicts better than
valence or dominance support for the President, levels of arousal are
also highly related to polarisation in valence: the higher the
heterogeneity in valence reactions, the higher are arousal levels.

The interaction of different emotional dimensions over time, and their
relationship to political attitudes, has gone largely unnoticed by
survey and poll research, but it has important repercussions for our
understanding of how the public think. The analyses above build a case
to start using online discussions and Internet-enabled communication
as sources of public opinion data. We have shown that online
discussions are representative of public opinion trends, even though
they are not demographically representative of the population. We have
also shown that the emotions identified in written communication can
be used as consistent indicators of political attitudes. As we already
mentioned above, this approach is not intended as a substitute for
other approaches to public opinion or emotions, but as an additional
tool to measure what is an elusive and complex dimension of human
behaviour. By analysing real-time reactions to political events, and
aggregating the emotional content of opinions voluntarily expressed,
we can overcome some of the measurement problems associated to survey
research; and by analysing large-scale, longitudinal data we can
overcome some of the limitations of experimental psychology, mostly
the threat of external validity.

Other implications of our research are theoretical. Research on public
opinion still misses an individual level mechanism that can explain
the aggregated shifts of opinions. Political psychology research has
focused on the cognitive effects of emotions, and on how they affect
information processing or settle predispositions for action; but it
does not offer empirical evidence on large-scale shifts in emotional
reactions, or the exact timing when reactions become more permanent
and manage to shift the mood of the public. By showing that emotional
reactions can predict approval rates we are strengthening the validity
of the measure, but also the theoretical explanation we can give to
changes in public opinion: these ultimately derive from individuals,
and emotions offer a crucial link between what individuals think and
what they do. Our findings generate some relevant questions in this
respect. We have found that arousal explains, more than valence,
support for the President, but it is a pending question whether that
is also the case for other expressions of public opinion like, say,
support for capital punishment or guns control. The ANEW words might
be missing important emotional dimensions that are crucially related
to political behaviour, so it is also a pending question whether
results would have been different if we had been able to identify
words implying anxiety as separate from words implying anger. The
arousal dimension mixes both types of emotions.  Finally, we also need
more research about the decaying rates of emotions as triggers for
action, and about whether those rates remain stable over time. Future
research should consider how emotional reactions are related to
behaviour, like voting, as opposed to just attitudes, and whether
their influence wanes with time.

These are all questions that require transcending the traditional ways
in which we have been measuring public opinion. In this paper, we have
provided a guideline on how to do that, profiting from the opinions
that the public voluntarily express in the forum created by online
discussions. Ours is a bottom-up approach that does not assume a list
of relevant issues about which we want the public to have an opinion,
but rather lets the public speak about the issues that most concern
them. This approach has a number of advantages: we analyse opinions on
a wider range of topics, we have less response bias, and we obtain
richer longitudinal information, involving significantly lower costs
in data collection than using sample surveys. Comparing how close the
measures of these two approaches are (i.e. the topics about which
people discuss voluntarily and the issues about which researchers have
been asking them for decades) is an interesting question in itself
that requires further investigation: we might find that not all the
``non-attitudes'' (Converse 1964) are an artefact of survey
methods. There are many alternative ways in which we can mine online
public opinion to measure trends in general sentiment (i.e. the
passions that drive political discourse). This paper has followed one
approach, but we need to investigate alternative ways to make emotions
operational concepts and find more fine-grained categories of
emotional response.

Moving beyond public opinion research, our results have also wider
implications for other areas of inquiry. There is an obvious marketing
edge in this approach, which is fitted to track public opinion about
products and goods, and project market revenues in line with what
customers (current or potential) think. Something along the lines has
been done by Asur and Huberman (2010); it would be interesting to
replicate their analysis with the ANEW lexicon and test if the
predictive power of the model changes significantly. There are also
implications for public health research: happiness has long been
acknowledged as an important component of health, and recent research
has found evidence of the contagious nature of emotions and their
ability to spread in the population (Fowler and Christakis
2008). Online discussion networks like those we analyse here might
also be channelling contagion processes; if so, this would offer a
crucial explanation of how public opinion is formed. Some other
studies have found that emotions are a crucial mechanism in the
dynamics of ``viral culture'': awe-inspiring newspaper articles, for
instance, are more likely to be among the most e-mailed stories on a
given day (Berger and Milkman 2010). If the news that inspire certain
emotional reactions are read by more people, then emotions are, again,
an important factor to explain the formation of opinions.

This takes us back to our findings: the same emotion-based viral
mechanism could be playing a role in the way people select topics from
their news sources, topics that they will then bring to the public
discussions we track. This agrees with political psychology research
on the way emotions can play a heuristic role in how citizens process
information. And this, in turn, highlights the most important moral of
this paper: that we need to explore further the explanatory power of
emotions if we are to distinguish alternative explanations of how
public opinion is formed. The analytical strategy we propose does not
allow us to make the usual demographic breakdowns (this information is
usually absent from digital data), but it sheds light into a
fundamental principle of human action, that is, the part of our
motivational structure that starts where rationality ends. Emotions
have been an elusive target for analysis on a large, societal scale;
we can move this line of research further by implementing new methods
that pay attention to the opinions that people voluntarily express.



\nocite{Adams1997}
\nocite{Adams1997}
\nocite{Asur2010}
\nocite{Baldassarri2008}
\nocite{Baldassarri2010}
\nocite{Berger2010}
\nocite{Box2008}
\nocite{Bradley1999}
\nocite{Carmines1989}
\nocite{Clarke2004}
\nocite{Converse1964}
\nocite{DelliCarpini1996}
\nocite{DiMaggio1996}
\nocite{Dodds2009}
\nocite{Elster1999}
\nocite{Elster2010}
\nocite{Erikson2002}
\nocite{Evans2003}
\nocite{Fiorina2005}
\nocite{Fisher2006}
\nocite{Fowler2008}
\nocite{Frank1988}
\nocite{Frijda1986}
\nocite{GonzalezBailonJIT2010}
\nocite{GonzalezBailon2010b}
\nocite{Glynn1999}
\nocite{Hauben1997}
\nocite{Huddy2007}
\nocite{Hutchings2005}
\nocite{Jacobs2000}
\nocite{Kriner2009}
\nocite{Layman2001}
\nocite{Lazarus1991}
\nocite{Lazarus1994}
\nocite{Lewis2001}
\nocite{Lippmann1922}
\nocite{Lueg2003}
\nocite{Marcus1993}
\nocite{Marcus2000}
\nocite{Mueller1973}
\nocite{Mutz2007}
\nocite{Neuman2007}
\nocite{OConnor2010}
\nocite{Page1992}
\nocite{Petrocik1987}
\nocite{Rahn2000}
\nocite{Redondo2007}
\nocite{Repass1971}
\nocite{Schuman1985}
\nocite{Smith2003}
\nocite{Smith1999}
\nocite{Sniderman1991}
\nocite{Stimson1998}
\nocite{Stimson2004}
\nocite{Stokes1963}
\nocite{Turner2006}
\nocite{Wlezien2005}
\nocite{Wolbrecht2000}
\nocite{Zaller1992}
\nocite{Zaller1992b}

\bibliographystyle{apacite}
\bibliography{references}
\newpage
\appendix
\section{Appendix}
\renewcommand{\thefigure}{\thesection.\arabic{figure}}  
\setcounter{figure}{0}
                    
\begin{figure}[!h]
\begin{center}
  \includegraphics[angle=-90,width=\textwidth]{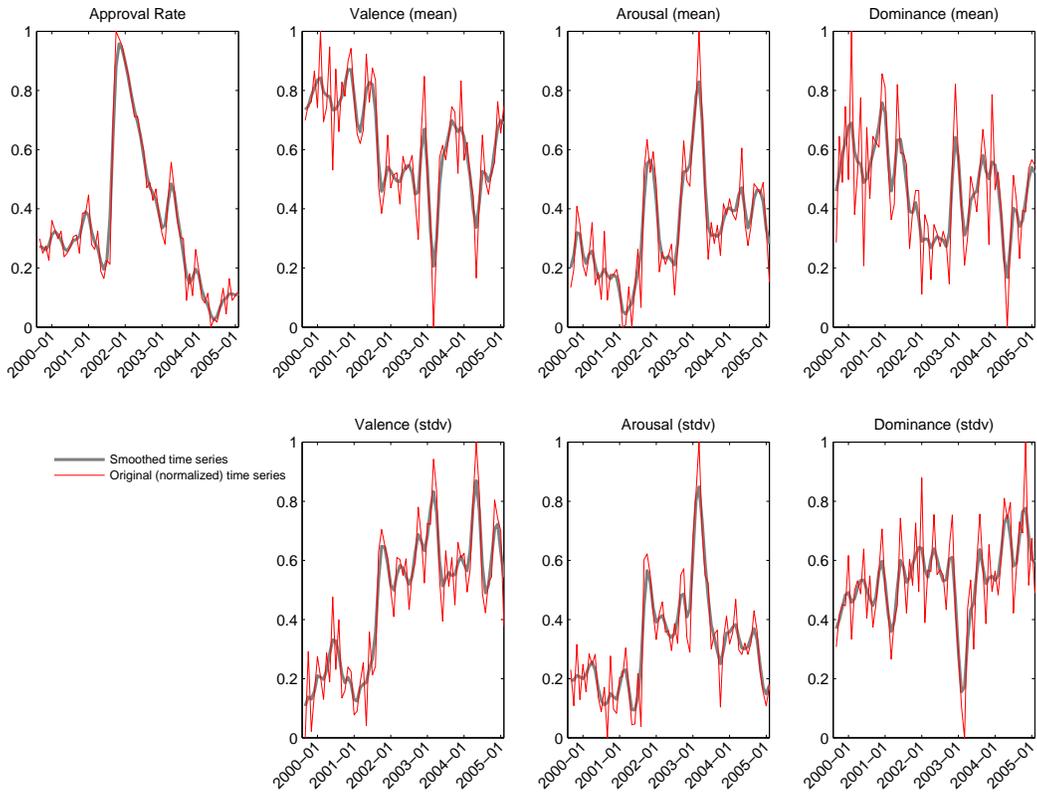}
\end{center}
\caption{Smoothed Time Series Used in the Analyses of Section 4.3.}
\label{fig:A1}
\end{figure}
\end{document}